\documentclass[10pt,final,conference,a4paper,oneside,twocolumn]{IEEEtran}
\IEEEoverridecommandlockouts

\usepackage{fixltx2e}

\usepackage{amssymb}
\usepackage[cmex10]{amsmath}
\usepackage{amstext}
\usepackage{array}
\usepackage{amsthm}
\usepackage{quoting}
\usepackage{tikz}
\usepackage{pgfplots}
\usepackage{import}
\usepackage{mathtools}
\usepgfplotslibrary{units}
\usetikzlibrary{spy,backgrounds}
\usepackage{pgfplotstable}
\usepackage{algorithm}
\usepackage{algorithmic}
\usepackage[mathscr]{euscript}

\usepackage[hidelinks,bookmarks=false]{hyperref}
\usepackage[figure]{hypcap}
\usepackage{indentfirst}
\usepackage{float}
\usepackage{cite}
\usepackage{microtype}
\usepackage[T1]{fontenc}
\usepackage[font=normalsize,labelfont=normalsize]{caption}
\usepackage{units}

\theoremstyle{definition}
\newtheorem{ass}{Assumption}[]

\newtheorem{theo}{Theorem}[]
\newtheorem{rem}{Remark}[]
\newlength\figureheight
\newlength\figurewidth

\makeatletter
\newcommand{\Eqref}[1]{\textup{\tagform@{\ref*{#1}}}}
\makeatother

\newcommand\copyrighttext{%
\Huge \textcopyright 2017 IEEE. Personal use of this material is permitted. Permission from IEEE must be obtained for all other uses, in any current or future media, including reprinting/republishing this material for advertising or promotional purposes, creating new collective works, for resale or redistribution to servers or lists, or reuse of any copyrighted component of this work in other works.}
\newcommand\copyrightnotice{%
\begin{tikzpicture}[remember picture,overlay]
\node[anchor=center] at (current page.center) {\fbox{\parbox{\dimexpr\textwidth-\fboxsep-\fboxrule\relax}{\copyrighttext}}};
\end{tikzpicture}%
}

\pgfplotsset{
compat=newest,
every axis/.append style={
font=\tiny
},
tick style={thin,black},
minor tick style={very thin,gray},
every axis legend/.append style={
draw=black,
fill=white,
shape=rectangle,
cells={anchor=center},
inner xsep=3pt,
inner ysep=2pt,
font=\tiny
}}

\usepackage{stackengine}
\usepackage{scalerel}
\usepackage{verbatimbox}
\newlength\glyphwidth
\newcommand\talldot{%
\ThisStyle{%
  \stackengine{.0ex}{$\SavedStyle.$}{$\SavedStyle.$}{O}{c}{F}{F}{S}%
}}
\newsavebox\hatglyphCONTENT
\sbox\hatglyphCONTENT{%
  \addvbuffer[-0.05ex -1ex]{%
    \stackengine{-.1ex}{ }%
                {\(\hat{}\)}{O}{c}{F}{F}{L}%
}}
\newcommand\hatglyph{\ThisStyle{\scalerel*{\usebox{\hatglyphCONTENT}}%
                     {\SavedStyle\talldot}}}
\newcommand\shifthat[2]{%
  \ThisStyle{%
    \stackengine{0.0ex}{\(\SavedStyle#2\)}%
                {\(\rule{#1}{0ex}\SavedStyle\hatglyph\)}{O}{c}{F}{T}{S}%
  }%
}

\usepackage[english]{babel}
\usepackage{blindtext}

\begin{document}

\copyrightnotice

\author{\IEEEauthorblockN{Bernardo Hernandez\IEEEauthorrefmark{1}$^{,a}$,
Paul Trodden\IEEEauthorrefmark{2}}
\IEEEauthorblockA{Department of Automatic Control and Systems Engineering, The University of Sheffield\\
Sheffield, S1 3JD, UK\\
Email: \IEEEauthorrefmark{1}bahernandezvicente1@sheffield.ac.uk,
\IEEEauthorrefmark{2}p.trodden@sheffield.ac.uk}
\thanks{$^{a}$Bernardo Hernandez acknowledges doctoral scholarship from CONICYT - PFCHA/Concurso para Beca de Doctorado en el Extranjero - 72150125, and financial support from the Dept. of Automatic Control \& Systems Engineering, Sheffield.}
}
\title{\LARGE \bf Distributed Model Predictive Control Using a Chain of Tubes}
\maketitle

\begin{abstract}
A new distributed MPC algorithm for the regulation of dynamically coupled subsystems is presented in this paper. At each time instant, the control action is computed via two robust controllers working in a nested fashion. The inner controller provides local reference trajectories computed on a fully decentralized framework. The outer controller uses this information to take into account the effects of the dynamic coupling and implement a distributed control action. The tube-based approach to robustness is employed. A supplementary constraint is included in the outer optimization problem to provide recursive feasibility of the overall controller.
\end{abstract}

\section{Introduction} \label{sec:Intro}

The current developments in fields such as wireless communication and actuation capabilities have led to an increase in the amount of information available to improve the task of controlling processes \begin{NoHyper}\cite{Ydstie2002}\end{NoHyper}. However, due to reasons such as limited computing capabilities and prohibitive cost of communication, standard centralized approaches are, sometimes, not well posed to take advantage of this additional information \begin{NoHyper}\cite{Christofides2013}\end{NoHyper}. Model predictive control (MPC) falls into the former. MPC is a mature control technique with guaranteed constraint satisfaction and closed loop stability \begin{NoHyper}\cite{Rawlings2014,Mayne2014}\end{NoHyper} (under proper assumptions and design), yet its implementation may turn infeasible for plants with a large number of inputs, given that it has to solve an optimization problem each time instant. Moreover, certain plants are naturally spread over a wide area, or arranged in an array of clearly defined subsystems, thus centralized control may not be the most practical choice \begin{NoHyper}\cite{Scattolini2009}\end{NoHyper}. A natural solution to this problem is to split the plant into smaller, easier to handle, subsystems, and then to synthesize local controllers. Depending on the type of plant being considered, and as a result of the division scheme, the subsystems may interact with each other directly through the dynamics of the global system, or through shared constraints; many non-centralized approaches have been presented to tackle these different configurations \begin{NoHyper}\cite{Scattolini2009,Christofides2013}\end{NoHyper}.

Subsystems subject to dynamic coupling are of particular interest. In order to provide any guarantee on the global system's behaviour, some coordination between agents must exist. In the context of linear time invariant (LTI) subsystems, cooperative \begin{NoHyper}\cite{Stewart2010,Venkat2006}\end{NoHyper} and non-cooperative \begin{NoHyper}\cite{Conte2013a,Wang2015a}\end{NoHyper} distributed controllers have been proposed. The latter ones are characterized, within the MPC framework, by locally defined cost functions that depend only on the particular subsystem variables. However, the different communication schemes allow the agents to obtain information about the future plans of its neighbours (hence, the coupling), thus a more informed choice of the current control action can be made. Known robust techniques, such as tube MPC \begin{NoHyper}\cite{Mayne2005}\end{NoHyper}, have been used to provide a system-wide feasible and stabilizing control solution to this type of system. In \begin{NoHyper}\cite{Farina2012a}\end{NoHyper}, reference trajectories are defined for each subsystem and the deviations from them are seen as a disturbance. Each subsystem is robustly controlled to reject these disturbances, and also to bound their own deviation. This approach is extended to output tracking in \begin{NoHyper}\cite{Farina2014}\end{NoHyper}, and to continuous time systems in \begin{NoHyper}\cite{Farina2014a}\end{NoHyper}. A similar scheme is presented in \begin{NoHyper}\cite{Riverso2012}\end{NoHyper}, where the tube approach is implemented twice, in series, for each subsystem.

In this paper, a new distributed non-cooperative MPC scheme is proposed for the regulation of dynamically coupled systems. At each time instant, the control action for each subsystem is computed locally through a two-step robust controller, however the inner step can be solved at a lower frequency. First, the whole dynamic interaction between neighbouring subsystems is regarded as an additive external disturbance, and a local tube-based controller is used to compute a feasible trajectory for each subsystem. These trajectories (state and input) are then regarded as references of what the subsystems will do, and shared amongst neighbours. A local outer controller, also based on the robust tube approach, is designed to take advantage of this information and further refine the predicted control trajectory, and the current control action for each subsystem. A supplementary constraint is enforced on the outer optimal control problem to guarantee recursive feasibility of the overall distributed controller and closed-loop asymptotic stability of the origin. The structure of the outer controller is similar to the one presented in \begin{NoHyper}\cite{Farina2012a,Farina2014a}\end{NoHyper}. The main difference is that the source of the reference trajectories is the inner decentralized controller, which means that no arbitrary sequence is needed at initialization and that the whole reference is allowed to change at each time step. This favours a reduction in conservativeness and an increase in performance. The inner-outer configuration is similar to the one presented in \begin{NoHyper}\cite{Riverso2012}\end{NoHyper} in that the tube approach is applied twice to compute the current control action by every agent. However in the scheme proposed in this paper, the tubes are applied in parallel (the control action of the outer controller is not a linear function of the inner controller result). The additional constraint enforced over the outer controller fulfils, essentially, the same role as the \emph{consistency} constraint in \begin{NoHyper}\cite{Dunbar2007}\end{NoHyper} or the \emph{reference tracking} constraints in \begin{NoHyper}\cite{Farina2012a}\end{NoHyper}, but it is constructed following invariance requirements for feasibility.

The paper is organized as follows: Section \ref*{sec:Pre} defines the problem and preliminaries. Section \ref*{sec:CTMPC} develops the proposed distributed controller. Recursive feasibility of the approach is analysed in Section \ref*{sec:RF} and some guidelines for design are discussed in Section \ref*{sec:CD}. Section \ref*{sec:SR} shows the behaviour of the proposed approach through a numerical example.

\emph{Notation}: At time $t$, $\hat{x}^{i}_{k/t}$ denotes the prediction of $\hat{x}^{i}$ at a future time $k$ within the horizon. The operators $\oplus$ and $\ominus$ denote the Minkowski sum and the Pontryagin difference (respectively), as defined in \begin{NoHyper}\cite{Rawlings2014}\end{NoHyper}. The zero vector and the identity matrix in $\mathbb{R}^n$ are, respectively, $0_n$ and $I_n$. $\left\vert\left\vert x\right\vert\right\vert^{2}_{Q}$ is the weighted squared norm of $x$. $\mathbb{X}^{\circ}$ represents the interior of the set $\mathbb{X}$. The symbol $\subset$ represent strict inclusion. The set $\mathscr{B}_{c_{n}}(\delta)\subset\mathbb{R}^{n}$ represents the ball of radius $\delta$ centred at $c_{n}$. The super index $^\top$ represents the transpose of a vector or matrix.

\section{Problem statement and preliminaries} \label{sec:Pre}
Consider the problem of regulating an LTI system for which a discrete time state space model is known,
\begin{equation} \label{eq:Pre.1}
\boldsymbol{x}_{t+1}=\boldsymbol{A}\boldsymbol{x}_{t}+\boldsymbol{B}\boldsymbol{u}_{t}
\end{equation}
where $\boldsymbol{x}\in\mathbb{R}^n$ is the state vector, $\boldsymbol{u}\in\mathbb{R}^m$ is the input vector and $\left(\boldsymbol{A},\boldsymbol{B}\right)$ are the state and input matrices of corresponding dimension. In many cases, it is convenient to represent \Eqref{eq:Pre.1} as a collection of $M$ coupled local subsystems. Suppose that \Eqref{eq:Pre.1} is arranged such that $\boldsymbol{x}=\left(x^{1\top},x^{2\top},\ldots,x^{M\top}\right)^{\top}$ represents a suitable non-overlapping decomposition of the state vector, and define $\mathcal{M}=\left\{1,\ldots,M\right\}$. The dynamics of each subsystem may be represented as follows,
\begin{subequations} \label{eq:Pre.2}
	\begin{alignat}{1}
	x^{i}_{t+1}&=A_{ii}x^{i}_{t}+B_{ii}u^{i}_{t}+\sum\nolimits_{j\in\mathcal{N}_i}\left(A_{ij}x^{j}_{t}+B_{ij}u^{j}_{t}\right)\label{eq:Pre.2.0}\\
	x^{i}_{t}&\in\mathbb{X}_{i},\:\: u^{i}_{t}\in\mathbb{U}_{i},\quad\forall t\geq 0\label{eq:Pre.2.1}
	\end{alignat}
\end{subequations}
where $x^{i}\in\mathbb{R}^{n_{i}}$ and $u^{i}\in\mathbb{R}^{m_{i}}$ are the subsystem state and input vectors, and $\left(A_{ij},B_{ij}\right)$ are blocks, of appropriate size, derived from the global matrices $\left(\boldsymbol{A},\boldsymbol{B}\right)$. The set $\mathcal{N}_i$ points to all the dynamic neighbours of subsystem $i$, i.e.:
\begin{equation} \label{eq:Pre.3}
\mathcal{N}_i=\left\{j\in\mathcal{M}\:|\:\left[A_{ij}\:B_{ij}\right]\neq 0\right\}
\end{equation}
\begin{ass}[Stabilizability] \label{ass:Pre.1}
The pairs $\left(\boldsymbol{A},\boldsymbol{B}\right)$ and $\left(A_{ii},B_{ii}\right)$ are stabilizable. Furthermore, there exists a block diagonal matrix $\boldsymbol{K}=diag\left(K_{11},\ldots,K_{MM}\right)$ such that $\boldsymbol{A}+\boldsymbol{B}\boldsymbol{K}$ and $A^{K_{ii}}_{ii}=A_{ii}+B_{ii}K_{ii}$ are Schur $\forall i\in\mathcal{M}$.
\end{ass}
\begin{rem}[] \label{rem:Pre.1}
Block-diagonal system-wide stabilizability is a standard (and required) assumption in the literature of distributed MPC controllers based on the robust Tube MPC technique \begin{NoHyper}\cite{Farina2012a,Farina2014,Farina2014a,Riverso2012}\end{NoHyper}.
\end{rem}
\begin{ass}[Properties of constraint sets] \label{ass:Pre.2}
For all $i\in\mathcal{M}$, the sets $\mathbb{X}_{i}$ and $\mathbb{U}_{i}$ are compact, convex and contain the origin in their interior.
\end{ass}

\section{Distributed control algorithm} \label{sec:CTMPC}
In this section, the distributed control algorithm is developed. At time $t$, the control action, for each subsystem, is obtained through a chain of two robust controllers. The outer controller employs the information produced by the inner one, to improve performance while rejecting disturbances generated by the dynamic coupling. Tube MPC -- a robust control technique designed to cope with bounded additive disturbances -- is employed for the inner and outer controllers. It solves the regulation problem for a nominal undisturbed model of the plant (subject to tightened constraints), while guaranteeing that the state of the true plant remains inside a robust positive invariant (RPI) set centred around the nominal trajectory.

\subsection{Decentralized reference definition}
The inner controller is designed on a decentralized framework, thus the whole dynamic interaction generated by the coupling with neighbouring subsystems is treated as an external disturbance, making tube MPC a suitable choice. The local dynamics in \Eqref{eq:Pre.2.0} can be simplified as,
\begin{subequations} \label{eq:DeTube.1}
	\begin{alignat}{2}
	x^{i}_{t+1}&=A_{ii}x^{i}_{t}+B_{ii}u^{i}_{t}+w^{i}_{t}&&\label{eq:DeTube.3}\\
	w^{i}_{t}\in\mathbb{W}_{i}&=\bigoplus_{j\in\mathcal{N}_i}\left(A_{ij}\mathbb{X}_{j}\oplus B_{ij}\mathbb{U}_{j}\right),&&\quad\forall t\geq 0 \label{eq:DeTube.2}
	\end{alignat}
\end{subequations}
which allows for a nominal undisturbed model to be defined:
\begin{subequations} \label{eq:DeTube.10}
	\begin{alignat}{2}
	\hat{x}^{i}_{t+1}&=A_{ii}\hat{x}^{i}_{t}+B_{ii}\hat{u}^{i}_{t}\label{eq:DeTube.4}\\
	\hat{x}^{i}_{t}&\in\hat{\mathbb{X}}_{i}\subseteq\mathbb{X}_{i}\ominus\mathbb{Z}_{i},&&\quad\forall t\geq 0 \label{eq:DeTube.10.1}\\
	\hat{u}^{i}_{t}&\in\hat{\mathbb{U}}_{i}\subseteq\mathbb{U}_{i}\ominus K_{T,i}\mathbb{Z}_{i}.&&\quad\forall t\geq 0  \label{eq:DeTube.10.2}
	\end{alignat}
\end{subequations}
$K_{T,i}$ is any stabilizing gain for the pair $\left(A_{ii},B_{ii}\right)$, guaranteed to exist in view of Assumption \ref*{ass:Pre.1}. The set $\mathbb{Z}_{i}$ is defined as an RPI set \begin{NoHyper}\cite{Kolmanovsky}\end{NoHyper} for the dynamics of $z^{i}_{t}=x^{i}_{t}-\hat{x}^{i}_{t}$, when the following disturbance rejection control policy is applied,
\begin{equation} \label{eq:DeTube.5}
u^{i}_{t}=\hat{u}^{i}_{t}+K_{T,i}\left(x^{i}_{t}-\hat{x}^{i}_{t}\right).
\end{equation}
At time $t$, the optimal nominal control action $\hat{u}^{i}_{t}$ is obtained from the optimization problem $\mathbb{P}_N(x^{i}_{t})$:
\begin{equation} \label{eq:DeTube.8}
\min_{\hat{x}^{i}_{t/t},\hat{u}^{i}_{[t:t+N-1]/t}}V_N\left(\hat{x}^{i}_{[t:t+N]/t},\hat{u}^{i}_{[t:t+N-1]/t}\right)
\end{equation}
subject to the dynamics in \Eqref{eq:DeTube.4} and:
\begin{subequations} \label{eq:DeTube.9}
	\begin{alignat}{1}
	x^{i}_{t}-\hat{x}^{i}_{t/t}&\in\mathbb{Z}_{i}, \label{eq:DeTube.9.1}\\
	\hat{x}^{i}_{k/t}&\in\hat{\mathbb{X}}_{i},\:\:\hat{u}^{i}_{k/t}\in\hat{\mathbb{U}}_{i},\quad k=t,\ldots,t+N-1 \label{eq:DeTube.9.3}\\
	\hat{x}^{i}_{t+N/t}&\in\hat{\mathbb{X}}^{F}_{i}. \label{eq:DeTube.9.5}
	\end{alignat}
\end{subequations}
The solution of \Eqref{eq:DeTube.8}--\Eqref{eq:DeTube.9} is a pair of optimal (nominal) input and state trajectories $\left(\hat{x}^{i*}_{[t:t+N/t]},\hat{u}^{i*}_{[t:t+N-1/t]}\right)$. Constraints \Eqref{eq:DeTube.9.1}--\Eqref{eq:DeTube.9.3} guarantee constraint satisfaction for the true plant in view of the invariance of the error dynamics, while the terminal constraint \Eqref{eq:DeTube.9.5} and the cost $V_N\left(\cdot\right)$ are designed to guarantee stability of the origin for the nominal system \begin{NoHyper}\cite{Rawlings2014}\end{NoHyper}.
\begin{ass}[Invariance of the inner terminal set] \label{ass:Pre.3}
The set $\hat{\mathbb{X}}^{F}_{i}$ is constraint admissible positive invariant for the closed-loop system $A^{K_{T,i}}_{ii}=A_{ii}+B_{ii}K_{T,i}$.
\end{ass}
Define the feasibility region of \Eqref{eq:DeTube.8} under \Eqref{eq:DeTube.9.3}--\Eqref{eq:DeTube.9.5} as $\hat{\mathcal{X}}^{i}_{N}$, if $x^{i}_{t}\in\hat{\mathcal{X}}^{i}_{N}\oplus\mathbb{Z}_{i}$, and \Eqref{eq:DeTube.5} is used to compute the true control action, then a feasible solution exists at time $t+1$, and the optimization is recursively feasible. However, in the nested approach presented here, \Eqref{eq:DeTube.5} is not used to compute the true control action. Instead, the optimized trajectories resulting from the inner controller are broadcast amongst neighbours.

\subsection{Distributed Tube MPC}
The outer controller regards the decentralized optimized trajectories as \emph{references} of what the subsystems will do, which allows to reduce the uncertainty about the dynamic coupling. In the following, the optimality super-index $\left(^*\right)$ is omitted. Define,
\begin{subequations} \label{eq:DTube.1}
	\begin{alignat}{1}
	d^{i}_{t/t}&=\sum_{j\in\mathcal{N}_i}\left(A_{ij}\hat{x}^{j}_{t/t}+B_{ij}\hat{u}^{j}_{t/t}\right) \label{eq:DTube.1.1}\\
	v^{i}_{t/t}&=\sum_{j\in\mathcal{N}_i}\left(A_{ij}\left(x^{j}_{t}-\hat{x}^{j}_{t/t}\right)+B_{ij}\left(u^{j}_{t/t}-\hat{u}^{j}_{t/t}\right)\right) \label{eq:DTube.1.2}
	\end{alignat}
\end{subequations}
where $d^{i}_{t/t}$ is known and $v^{i}_{t/t}\in\mathbb{V}_{i}$. This allows to recast \Eqref{eq:DeTube.3} as,
\begin{equation} \label{eq:DTube.2}
x^{i}_{t+1}=A_{ii}x^{i}_{t}+B_{ii}u^{i}_{t}+d^{i}_{t/t}+v^{i}_{t/t}
\end{equation}
where $v^{i}_{t/t}$ is a bounded additive disturbance that represents the deviation of the true plant trajectory, from the decentralized nominal optimal trajectory. The objective of the outer controller is to reduce the uncertainty about the dynamic coupling, therefore the following assumption is in order:
\begin{ass}[Uncertainty reduction] \label{ass:CTMC.1}
$\mathbb{V}_{i}\subset\mathbb{W}^{\circ}_{i},\:\forall i\in\mathcal{M}$
\end{ass}
The outer controller is also synthesized as a tube MPC. In this case, the nominal model takes the following form,
\begin{subequations} \label{eq:DTube.4}
	\begin{alignat}{2}
	\hat{\hat{x}}^{i}_{t+1}&=A_{ii}\hat{\hat{x}}^{i}_{t}+B_{ii}\hat{\hat{u}}^{i}_{t}+d^{i}_{t/t}, \label{eq:DTube.4.1}\\
	\hat{\hat{x}}^{i}_{k/t}&\in\hat{\hat{\mathbb{X}}}_{i}\subseteq\mathbb{X}_{i}\ominus\mathbb{S}_{i},&&\quad\forall t\geq 0 \label{eq:DTube.10.1}\\
	\hat{\hat{u}}^{i}_{k/t}&\in\hat{\hat{\mathbb{U}}}_{i}\subseteq\mathbb{U}_{i}\ominus \hat{K}_{i}\mathbb{S}_{i},&&\quad\forall t\geq 0 \label{eq:DTube.10.2}
	\end{alignat}
\end{subequations}
where $\hat{K}_{i}$ is any stabilizing gain for the pair $\left(A_{ii},B_{ii}\right)$, guaranteed to exist in view of Assumption \ref*{ass:Pre.1}. The set $\mathbb{S}_{i}$ is defined as an RPI set for the dynamics of $s^{i}_{t}=x^{i}_{t}-\hat{\hat{x}}^{i}_{t}$, when the following disturbance rejection control law is applied,
\begin{equation} \label{eq:DTube.5}
u^{i}_{t}=\hat{\hat{u}}^{i}_{t}+\hat{K}_{i}\left(x^{i}_{t}-\hat{\hat{x}}^{i}_{t}\right).
\end{equation}

At time $t$, the optimal nominal control action $\hat{\hat{u}}^{i}_{t}$ is obtained from the optimization problem $\mathbb{P}^{2}_N(x^{i}_{t})$:
\begin{equation} \label{eq:DTube.8}
\min_{\hat{\hat{x}}^{i}_{t/t},\hat{\hat{u}}^{i}_{[t:t+N-1]/t}}\hat{V}_N\left(\hat{\hat{x}}^{i}_{[t:t+N]/t},\hat{\hat{u}}^{i}_{[t:t+N-1]/t}\right)
\end{equation}
subject to the dynamics in \Eqref{eq:DTube.4.1} and:
\begin{subequations} \label{eq:DTube.9}
	\begin{alignat}{2}
	x^{i}_{t}-\hat{\hat{x}}^{i}_{t/t}&\in\mathbb{S}_{i} \label{eq:DTube.9.1}\\
	\hat{\hat{x}}^{i}_{k/t}&\in\hat{\hat{\mathbb{X}}}_{i},\;\;\hat{\hat{u}}^{i}_{k/t}\in\hat{\hat{\mathbb{U}}}_{i},\,&& k=t,\ldots,t+N-1 \label{eq:DTube.9.3}\\
	\hat{\hat{x}}^{i}_{t+N/t}&\in\hat{\hat{\mathbb{X}}}^{F}_{i} \label{eq:DTube.9.5}\\
	\hat{\hat{x}}^{i}_{k/t}-\hat{x}^{i}_{k/t}&\in\mathbb{H}_{i}.\quad&& k=t+1,\ldots,t+N \label{eq:DTube.9.6}
	\end{alignat}
\end{subequations}
where the terminal constraint sets must fulfil the following assumption:
\begin{ass}[Invariance of the outer terminal set] \label{ass:Pre.4}
The set $\hat{\hat{\mathbb{X}}}^{F}=\prod_{i=1}^{M}\hat{\hat{\mathbb{X}}}^{F}_{i}$ is a constraint admissible positive invariant set for the closed loop dynamics $\left(\boldsymbol{A}+\boldsymbol{B}\boldsymbol{\hat{K}}\right)$.
\end{ass}
The structure of the optimization problem \Eqref{eq:DTube.8}--\Eqref{eq:DTube.9} differs from a standard tube MPC only in the addition of constraint \Eqref{eq:DTube.9.6}. This supplementary constraint has two purposes: guarantee recursive feasibility of the overall controller, and provide a comprehensive way for computing the set $\mathbb{V}_{i}$. The design of the set $\mathbb{H}_{i}$ is discussed later.

\subsection{Control Algorithm}
The proposed controller is summarized in Algorithm \ref*{alg:1}.
 \begin{algorithm}[H]
 \caption{} \label{alg:1}
 \begin{algorithmic}[1]
 \renewcommand{\algorithmicrequire}{\textbf{Input:}}
\STATE $\lambda=0$
\STATE Measure $\boldsymbol{x}_{t}$ \label{alg:step.0}
\FOR {$i\in\mathcal{M}$}
\IF {$t=\lambda T$} \label{alg:step.1}
\STATE Solve \Eqref{eq:DeTube.8} subject to \Eqref{eq:DeTube.9}.
\STATE $\lambda=\lambda+1$
\ELSE
\STATE $\hat{x}^{i}_{[t:t+N-1]/t}=\left[\hat{x}^{i}_{[t:t+N-1]/t-1}\:\: A^{K_{T,i}}_{ii}\hat{x}^{i}_{t+N-1/t-1}\right]$
\STATE $\hat{u}^{i}_{[t:t+N-1]/t}=\left[\hat{u}^{i}_{[t:t+N-2]/t-1}\:\: K_{T,i}\hat{x}^{i}_{t+N-1/t-1}\right]$ \label{alg:step.2}
\ENDIF
\ENDFOR
\STATE Broadcast the optimal solution provided by the inner controller $\left(\hat{x}^{i}_{[t:t+N/t]},\hat{u}^{i}_{[t:t+N-1/t]}\right)$ amongst neighbours.
\FOR {$i\in\mathcal{M}$}
\STATE Solve \Eqref{eq:DTube.8} subject to \Eqref{eq:DTube.9}.
\STATE Compute $u^{i}_{t}=\hat{\hat{u}}^{i}_{t/t}+\hat{K}_{i}\left(x^{i}_{t}-\hat{\hat{x}}^{i}_{t/t}\right)$ and apply to true plant.
\ENDFOR
\STATE set $t=t+1$ and go to step \ref*{alg:step.0}.
\end{algorithmic}
\end{algorithm}
Steps \ref*{alg:step.1}--\ref*{alg:step.2} of Algorithm \ref*{alg:1} define how the reference trajectories are updated. Each $T$ time instants, the optimization \Eqref{eq:DeTube.8}--\Eqref{eq:DeTube.9} is solved and the trajectories are updated in its whole. At any other time instant, the reference trajectories are updated making use of the local terminal controller defined by $K_{T,i}$.

\section{Feasibility of the distributed controller} \label{sec:RF}
This section analyses the recursive feasibility of the distributed controller through the design of the sets $\mathbb{H}_{i}$. The following results consider the controllers as they have been presented up to now, in their most general form. A set of simple design choices are presented in Section \ref*{sec:CD}.
\subsection{Backwards recursive feasibility} \label{sec:RF.1}
Satisfaction of the RPI constraint \Eqref{eq:DeTube.9.1} guarantees $z^{i}_{t/t}\in\mathbb{Z}_{i}$ however, the control policy in use is \Eqref{eq:DTube.5} (not \Eqref{eq:DeTube.5}) hence,
\begin{equation} \label{eq:DTube.12}
z^{i}_{t+1/t}=A_{ii}z^{i}_{t/t}+B_{ii}\left(u^{i}_{t}-\hat{u}^{i}_{t/t}\right)+w^{i}_{t/t}
\end{equation}
which is not necessarily inside $\mathbb{Z}_{i}$. It follows then, that recursive feasibility of the inner nominal problem could be broken. Constraint \Eqref{eq:DTube.9.6}, called hereafter the backwards recursive feasibility (BRF) constraint, is designed to account for this effect. Consider the following result.
\begin{theo}[RPI sets inclusion] \label{theo:RF.1}
If Assumption \ref*{ass:CTMC.1} holds, then there exists a pair of linear gains $\left(K_{T,i},\hat{K}_{i}\right)$ such that $\mathbb{S}_{i}\subset\mathbb{Z}^{\circ}_{i}$.
\end{theo}
\begin{proof}
The proof is omitted for brevity.
\end{proof}

In practice, theorem \ref*{theo:RF.1} allows for the existence of a positive scalar $\delta_{i}$ such that $\mathbb{S}_{i}\oplus\mathscr{B}_{0_{n_{i}}}(\delta_{i})\subseteq\mathbb{Z}_{i}$. Suppose from now on $\mathbb{H}_{i}\subseteq\mathscr{B}_{0_{n_{i}}}(\delta_{i})$, then the following is easily derived:
\begin{subequations} \label{eq:DTube.26}
	\begin{alignat}{1}
	&\left(\hat{\hat{x}}^{i}_{t+1/t}-\hat{x}^{i}_{t+1/t}\right)\in\mathbb{H}_{i}\\
	&\implies z^{i}_{t+1/t}\in\mathbb{S}_{i}\oplus\mathbb{H}_{i}\subseteq\mathbb{Z}_{i}
	\end{alignat}
\end{subequations}
Hence, a proper selection of the linear gains, such that $\mathbb{S}_{i}\subset\mathbb{Z}^{\circ}_{i}$, guarantees BRF.

\subsection{Recursive feasibility} \label{sec:RF.2}
Define the feasibility region of \Eqref{eq:DTube.8} under \Eqref{eq:DTube.9.3}--\Eqref{eq:DTube.9.5} as $\hat{\hat{\mathcal{X}}}^{i}_{N}$, if $x^{i}_{t}\in\hat{\hat{\mathcal{X}}}^{i}_{N}\oplus\mathbb{S}_{i}$, then a feasible solution exists at time $t+1$, and the optimization is recursively feasible. To provide a guarantee of recursive feasibility when the BRF constraint \Eqref{eq:DTube.9.6} is enforced, define $e^{i}_{t}=\hat{\hat{x}}^{i}_{t}-\hat{x}^{i}_{t}$ and consider the following design assumption,
\begin{ass}[$\mathbb{H}_{i}$ design condition] \label{ass:CTMC.2}
The sets $\mathbb{H}_{i}$ are designed such that,
\begin{subequations}
	\begin{alignat}{1}
	\left(A_{ii}+B_{ii}\hat{K}_{i}\right)\mathbb{H}_{i}&\oplus\left(B_{ii}\left(\hat{K}_{i}-K_{T,i}\right)\hat{\hat{\mathbb{X}}}^{F}_{i}\right)\oplus\mathbb{D}^{F}_{i}\subseteq\mathbb{H}_{i} \label{eq:RF.1}\\
	\mathbb{D}^{F}_{i}&=\bigoplus_{j\in\mathcal{N}_i}\left(A_{ij}+B_{ij}K_{T,j}\right)\hat{\mathbb{X}}^{F}_{j}
	\end{alignat}
\end{subequations}
\end{ass}

\begin{theo}[Recursive feasibility] \label{theo:RF.2}
If Assumptions \ref*{ass:Pre.3}, \ref*{ass:Pre.4} and \ref*{ass:CTMC.2} hold, and a feasible solution exists for problems \Eqref{eq:DeTube.8}--\Eqref{eq:DeTube.9} and \Eqref{eq:DTube.8}--\Eqref{eq:DTube.9} at time $t$, then a feasible solution exists, for both optimization problems, at time $t+1$.
\end{theo}
\begin{proof}
The proof is omitted for brevity.
\end{proof}

\subsection{Feasibility of the distributed controller} \label{sec:RF.3}
Up to this point, the recursive feasibility feature has been addressed separately for both optimization problems. However, it is important to note that constraint \Eqref{eq:DTube.9.6} is parametrised by the result of the inner optimization, hence the overall feasibility region of the distributed controller could be small. To address this issue, consider the following result,
\begin{theo}[Recursive feasibility of the overall controller] \label{theo:RF.3}
If constraint \Eqref{eq:DeTube.9.1} is replaced by,
\begin{equation}
x^{i}_{t}-\hat{x}^{i}_{t/t}\in\mathbb{S}_{i}\oplus\mathbb{H}_{i},
\end{equation}
the set $\mathbb{D}^{F}_{i}$ in Assumption \ref*{ass:CTMC.2} is replaced by,
\begin{equation}
\mathbb{D}_{i}=\bigoplus_{j\in\mathcal{N}_i}\left(A_{ij}\hat{\mathbb{X}}_{j}\oplus B_{ij}\hat{\mathbb{U}}_{j}\right)
\end{equation}
and the terminal sets in \eqref{eq:DeTube.9.5} are designed such that
\begin{equation}
\hat{\mathbb{X}}^{F}_{i}\subseteq\hat{\hat{\mathbb{X}}}^{F}_{i}\ominus\mathbb{H}_{i},
\end{equation}
then a feasible solution of the optimization problem \Eqref{eq:DeTube.8}--\Eqref{eq:DeTube.9} at time $t$, implies that,
\begin{equation}
\hat{\hat{u}}^{i}_{k/t}=\hat{u}^{i}_{k/t}+\hat{K}_{i}\left(\hat{\hat{x}}^{i}_{k/t}-\hat{x}^{i}_{k/t}\right),\quad k=t,\ldots,t+N-1
\end{equation}
is a feasible solution for the problem \Eqref{eq:DTube.8}--\Eqref{eq:DTube.9} at time $t$.
\end{theo}
\begin{proof}
The proof is omitted for brevity.
\end{proof}

\subsection{Disturbance computation} \label{sec:RF.4}
The second purpose of the additional constraint \Eqref{eq:DTube.9.6} is to provide a comprehensive way of computing the set $\mathbb{V}_{i}$. To define $\mathbb{V}_{i}$ it is necessary to bound the deviation between the reference and true trajectories for both, state and input \Eqref{eq:DTube.1.2}. Satisfaction of the RBF constraint \Eqref{eq:DTube.9.6} already bounds the state deviation to lie inside $\mathbb{Z}_{i}$ throughout the horizon \Eqref{eq:DTube.26}
\begin{equation} \label{eq:RF.2}
x^{i}_{k/t}-\hat{x}^{i}_{k/t}\in\mathbb{Z}_{i}.
\end{equation}
Now let the BRF constraint be restated as,
\begin{equation}
e^{i}_{k/t}\in\mathbb{H}_{i},\quad k=t+1,\ldots,t+N
\end{equation}
and define the dynamics of the error between both nominal trajectories (throughout the horizon) as follows:
\begin{equation}
e^{i}_{k+1/t}=A_{ii}e^{i}_{k/t}+B_{ii}\left(\hat{\hat{u}}^{i}_{k/t}-\hat{u}^{i}_{k/t}\right)+d^{i}_{k/t}.
\end{equation}
Clearly then, satisfaction of the BRF constraint \Eqref{eq:DTube.9.6} implies an invariance inducing behaviour of $\left(\hat{\hat{u}}^{i}_{k/t}-\hat{u}^{i}_{k/t}\right)$, which in turn means that $\left(\hat{\hat{u}}^{i}_{k/t}-\hat{u}^{i}_{k/t}\right)$ must be bounded. Denominate the set that bounds the input deviation as $\shifthat{-0.31em}{\mathbb{L}}_{i}$, it follows that,
\begin{subequations} \label{eq:DTube.27}
	\begin{alignat}{2}
	&\left(\hat{\hat{u}}^{i}_{k/t}-\hat{u}^{i}_{k/t}\right)&&\in\shifthat{-0.31em}{\mathbb{L}}_{i}\\
	\implies&\left(u^{i}_{k/t}-\hat{u}^{i}_{k/t}\right)&&\in\hat{K}_{i}\mathbb{S}_{i}\oplus\shifthat{-0.31em}{\mathbb{L}}_{i}\subseteq\mathbb{L}_{i} \label{eq:RF.3}
	\end{alignat}
\end{subequations}
The specific computation of the set $\shifthat{-0.31em}{\mathbb{L}}_{i}$ is out of the scope of this paper, hence the definition of the auxiliary set $\mathbb{L}_{i}$. In view of \Eqref{eq:RF.2} and \Eqref{eq:RF.3}, the disturbance sets $\mathbb{V}_{i}$ can be defined as,
\begin{equation}
\mathbb{V}_{i}=\bigoplus_{j\in\mathcal{N}_i}\left(A_{ij}\mathbb{Z}_{j}+B_{ij}\mathbb{L}_{j}\right)
\end{equation}

\section{Controller design} \label{sec:CD}
Constraint \Eqref{eq:DTube.9.6} adds complexity in the construction of the controllers. However, a simple approach can be taken in the selection of the key design parameters.

\subsection{Linear gains for recursive feasibility}
Suppose that Assumption \ref*{ass:CTMC.1} is fulfilled by the trivial choice, $\mathbb{L}_{i}=\mathbb{U}_{i}+\bar{\hat{\mathbb{U}}}_{i}$ (where $\bar{\hat{\mathbb{U}}}_{i}$ is a linear map of the set $\hat{\mathbb{U}}_{i}$ by the transformation matrix $-I_{n_{i}}$). Clearly then, setting $\hat{K}_{i}=K_{T,i}$ guarantees $\mathbb{S}_{i}\subset\mathbb{Z}^{\circ}_{i}$. The trivial choice of the linear gains also shortens \Eqref{eq:RF.1} to,
\begin{equation} \label{eq:RF.40}
\left(A_{ii}+B_{ii}\hat{K}_{i}\right)\mathbb{H}_{i}\oplus\mathbb{D}^{F}_{i}\subseteq\mathbb{H}_{i}
\end{equation}
which means that the set $\mathbb{H}_{i}\subset\mathbb{Z}_{i}\cap\mathbb{S}^{c}_{i}$ must be computed as an RPI set for the closed loop dynamics $\left(A_{ii}+B_{ii}\hat{K}_{i}\right)$ under the effect of a disturbance contained in $\mathbb{D}^{F}_{i}$. 

\subsection{Cost function for stability}
A standard quadratic cost function with terminal state penalty is chosen for the outer controller,
\begin{equation}
\begin{alignedat}{1} \label{eq:D.11}
& \hat{V}_{N}\left(\hat{\hat{x}}^{i}_{[t:t+N]/t},\hat{\hat{u}}^{i}_{[t:t+N-1]/t}\right)=\\
&\sum^{t+N-1}_{k=t}\left(\left\vert\left\vert \hat{\hat{x}}^{i}_{k/t}\right\vert\right\vert^{2}_{Q_{i}}+\left\vert\left\vert \hat{\hat{u}}^{i}_{k/t}\right\vert\right\vert^{2}_{R_{i}}\right)+\left\vert\left\vert \hat{\hat{x}}^{i}_{t+N/t}\right\vert\right\vert^{2}_{P_{i}}
\end{alignedat}
\end{equation}
The weight matrices $Q_{i}$, $R_{i}$ and $P_{i}$ are built following the procedure depicted in \begin{NoHyper}\cite{Farina2012a}\end{NoHyper} to guarantee global stability. For comparison purposes and to facilitate the design, the same cost function is used for the inner controller.

\begin{theo}[Asymptotic stability of the closed-loop system] \label{theo:CD.1}
If the cost functions $V_{N}$ and $\hat{V}_{N}$ are designed as in \Eqref{eq:D.11}, and the terminal sets following Assumptions \ref*{ass:Pre.2} and \ref*{ass:Pre.3}, then the set $\mathcal{A}\coloneqq\left\{\boldsymbol{0}_{n_{i}}\right\}\times\left\{\boldsymbol{0}_{n_{i}}\right\}\times\left\{\boldsymbol{0}_{n_{i}}\right\}$ is asymptotically stable for the constrained composite system,
\begin{equation*}
	\begin{aligned}
	x^{i}_{t+1}&=A_{ii}x^{i}_{t}+B_{ii}u^{i}_{t}+w^{i}_{t}\\
	\hat{x}^{i}_{t+1}&=A_{ii}\hat{x}^{i}_{t}+B_{ii}\hat{u}^{i}_{t}\\
	\hat{\hat{x}}^{i}_{t+1}&=A_{ii}\hat{\hat{x}}^{i}_{t}+B_{ii}\hat{\hat{u}}^{i}_{t}+d^{i}_{t}
	\end{aligned}
\end{equation*}
under the closed-loop control laws defined by \Eqref{eq:DTube.5}, \Eqref{eq:DeTube.8}--\Eqref{eq:DeTube.9}, and \Eqref{eq:DTube.8}--\Eqref{eq:DTube.9} respectively.
\end{theo}
\begin{proof}
The proof is omitted for brevity.
\end{proof}

\section{Simulation results} \label{sec:SR}
A slightly modified version of the four truck system depicted in \begin{NoHyper}\cite{Riverso2012}\end{NoHyper} is used in this section to show the behaviour of the proposed control algorithm. The plant consists of four trucks represented by point masses and dynamically coupled through springs and dampers; each one is linked only to its immediate neighbours. The control objective is to steer the whole system towards the origin. Note that this plant is marginally stable when uncontrolled.

The global system is composed of eight state variables and 4 input variables. A smaller subsystem is associated to each truck, $M=4$, with two state variables representing its horizontal position and velocity (w.r.t. an arbitrary equilibrium point) and one input variable, representing an horizontal force applied to the truck. The masses are: $m_{1}=3$, $m_{2}=2$, $m_{3}=3$, $m_{4}=6$; the spring constants are $k_{12}=7.5$, $k_{23}=0.75$, $k_{34}=1$; the damper coefficients are $c_{12}=4$, $c_{23}=0.25$, $c_{34}=0.3$. The coupling between trucks 1 and 2 is purposely designed higher than the rest, in order to increase the size of $\mathbb{W}_{1,2}$ and show the behaviour of the proposed approach under high and low couplings. Each truck is subject to the following state and input constraints:
\begin{subequations} \label{eq:SR.3}
	\begin{alignat}{3}
		\mathbb{X}_{i}&=\left\{x^{i}\:\Big\vert\:\begin{bmatrix}
		-2 \\
		-8
		\end{bmatrix}\leq x^{i} \leq \begin{bmatrix}
		2 \\
		8
		\end{bmatrix}\right\},\quad&&i=1,2,3,4\\
		\mathbb{U}_{i}&=\left\{u^{i}\:\vert\:-4\leq u^{i}\leq 4\right\},\quad&&i=1,2,3,4
	\end{alignat}
\end{subequations}

The distributed controller is designed following the approach described in Sections \ref{sec:RF} and \ref{sec:CD}. A sampling time of $T_{s}=0.1\left[s\right]$ is used to discretize the system. For brevity, the values of each design parameter are not depicted in this paper. The discretised system is initialised at a point that is feasible but close to the feasibility boundaries,
\begin{equation} \label{eq:SR.6}
x^{1}_{0}=\begin{bmatrix*}[r]1.8\\-2.0\end{bmatrix*},x^{2}_{0}=\begin{bmatrix*}[r]0.5\\7.1\end{bmatrix*} x^{3}_{0}=\begin{bmatrix*}[r]-0.9\\-7.0\end{bmatrix*},x^{4}_{0}=\begin{bmatrix*}[r]-1.8\\2.0\end{bmatrix*}
\end{equation}

\begin{figure}[!t]
	\centering
	\pgfplotsset{major grid style={dashed,gray}}
	\pgfplotsset{minor grid style={dotted,gray}}
	\pgfplotsset{
		every axis legend/.append style={
		at={(0.95,0.94)},
		anchor=north east,
		row sep=-3pt,
	}}
	\setlength\figureheight{0.2\textwidth}
	\setlength\figurewidth{0.4\textwidth}
	
	\definecolor{mycolor1}{rgb}{1.00000,0.00000,1.00000}%
	\begin{tikzpicture}
	
	\begin{axis}[%
	width=\figurewidth,
	height=\figureheight,
	at={(0\figurewidth,0\figureheight)},
	scale only axis,
	xmin=-2,
	xmax=2,
	xlabel={Truck position},
	xtick={-2,-1.5,...,2},
	xticklabel style = {font=\scriptsize},
	xlabel style = {font=\scriptsize},
	xmajorgrids,
	xminorgrids,
	ymin=-8,
	ymax=8,
	ylabel={Truck velocity},
	ytick={-8,-6,...,8},
	yticklabel style = {font=\scriptsize},
	ylabel style = {font=\scriptsize},
	ymajorgrids,
	yminorgrids,
	legend style={legend plot pos=left},
	scaled ticks = false,
	mark repeat={3}
	]
	
	\addplot [color=black,solid,mark=asterisk,mark options={solid}]
	  table[row sep=crcr]{%
	   1.800000000000000  -2.000000000000000\\
	   1.465615006851268  -4.615609597664649\\
	   1.013007538158220  -4.421983967822482\\
	   0.631730368991285  -3.214517949674889\\
	   0.369349807922522  -2.048093893198891\\
	   0.206719576872104  -1.216229460490498\\
	   0.110482426183950  -0.715633520190633\\
	   0.053963841286344  -0.418978821664114\\
	   0.022926594706335  -0.205225364438006\\
	   0.008603307854417  -0.083353221955169\\
	   0.003056273871331  -0.028574267544019\\
	   0.001225794949702  -0.008407424578358\\
	   0.000679494696744  -0.002624396016333\\
	   0.000476784639522  -0.001447524607599\\
	   0.000342579540825  -0.001235694675596\\
	   0.000229520228288  -0.001025454713129\\
	   0.000141342673600  -0.000740751999862\\
	   0.000080784322009  -0.000473826331989\\
	   0.000043509079788  -0.000274555803454\\
	   0.000022494654211  -0.000147678022632\\
	   0.000011395374259  -0.000075450958915\\
	   0.000005741370364  -0.000038221277848\\
	   0.000003010915686  -0.000016751397837\\
	   0.000001731181595  -0.000008963121114\\
	   0.000001058438886  -0.000004561563683\\
	   0.000000733159947  -0.000001987688144\\
	   0.000000578665816  -0.000001115156445\\
	   0.000000489719808  -0.000000669955858\\
	   0.000000438660660  -0.000000356078170\\
	   0.000000414684384  -0.000000127529802\\
	   0.000000402204408  -0.000000121679734\\
	   0.000000394805381  -0.000000028073719\\
	   0.000000390100513  -0.000000065006421\\
	   0.000000385152464  -0.000000034428247\\
	   0.000000382991236  -0.000000009265708\\
	   0.000000382609012   0.000000001399995\\
	   0.000000383117323   0.000000008585300\\
	   0.000000383499800  -0.000000000744279\\
	   0.000000383426998  -0.000000000709488\\
	   0.000000383367486  -0.000000000483301\\
	   0.000000383330349  -0.000000000262794\\
	   0.000000383283305  -0.000000000667145\\
	   0.000000383249698  -0.000000000018021\\
	   0.000000383516408   0.000000005224547\\
	   0.000000383627109  -0.000000002835931\\
	   0.000000383375100  -0.000000002207706\\
	   0.000000383225606  -0.000000000807100\\
	   0.000000383181799  -0.000000000083305\\
	   0.000000383179401   0.000000000032879\\
	   0.000000383174682  -0.000000000123587\\
	   0.000000383457867   0.000000005647170\\
	}; \addlegendentry{$x^{1}$}
	\addplot [color=mycolor1,solid,mark=o,mark options={solid}]
	  table[row sep=crcr]{%
	   0.508700000000000   7.100000000000000\\
	   0.745370526287580  -2.058301163749777\\
	   0.496670477301211  -2.869842827290197\\
	   0.255982644324291  -1.959624589258298\\
	   0.108163530206470  -1.019898475105127\\
	   0.033175626218838  -0.493364218144307\\
	  -0.000615583368132  -0.190857429871770\\
	  -0.011691496720918  -0.035408333039110\\
	  -0.012597397765000   0.015546804635480\\
	  -0.010067426832124   0.034217040964858\\
	  -0.006716975049540   0.032618322794532\\
	  -0.003939838908990   0.023071800856715\\
	  -0.002107240317652   0.013783358485732\\
	  -0.001059897116475   0.007320539376548\\
	  -0.000518725364810   0.003597627718263\\
	  -0.000255981617027   0.001706214046143\\
	  -0.000131674024364   0.000803387236224\\
	  -0.000072047634366   0.000399359731096\\
	  -0.000041509798493   0.000215802672903\\
	  -0.000024404845767   0.000128224162573\\
	  -0.000014287379819   0.000075305012575\\
	  -0.000008320690843   0.000044703077067\\
	  -0.000004787963875   0.000026358704128\\
	  -0.000002677308347   0.000016073792371\\
	  -0.000001464525949   0.000008371443569\\
	  -0.000000839411454   0.000004234481127\\
	  -0.000000545706548   0.000001708928425\\
	  -0.000000410810532   0.000001004676191\\
	  -0.000000303717878   0.000001125531783\\
	  -0.000000246086580   0.000000060862794\\
	  -0.000000219246894   0.000000459942925\\
	  -0.000000210782498  -0.000000265527165\\
	  -0.000000229865635  -0.000000119981637\\
	  -0.000000236726018  -0.000000020287450\\
	  -0.000000236692309   0.000000019547615\\
	  -0.000000231858881   0.000000074829548\\
	  -0.000000228600377  -0.000000006984096\\
	  -0.000000228901316   0.000000000713213\\
	  -0.000000228844883   0.000000000421901\\
	  -0.000000228784258   0.000000000773945\\
	  -0.000000228813852  -0.000000001290631\\
	  -0.000000228806393   0.000000001345867\\
	  -0.000000226122950   0.000000050400102\\
	  -0.000000225006741  -0.000000025464120\\
	  -0.000000226958283  -0.000000013844358\\
	  -0.000000227922360  -0.000000005660902\\
	  -0.000000228334582  -0.000000002661506\\
	  -0.000000228522128  -0.000000001130772\\
	  -0.000000228617567  -0.000000000783734\\
	  -0.000000226037709   0.000000050390008\\
	  -0.000000225002981  -0.000000027023161\\
	}; \addlegendentry{$x^{2}$}
	\addplot [color=red,solid,mark=diamond,mark options={solid}]
	  table[row sep=crcr]{%
	  -0.907700000000000  -7.000000000000000\\
	  -1.057864895307967   3.964652730465833\\
	  -0.669891635001367   3.791569134218735\\
	  -0.365506502040563   2.297739296128228\\
	  -0.189174348214153   1.230451057402783\\
	  -0.095919256459080   0.635562107462069\\
	  -0.047987705990931   0.323556388891050\\
	  -0.023623851786343   0.163971515240130\\
	  -0.011418271577798   0.080277193646246\\
	  -0.005420644036320   0.039740925785791\\
	  -0.002483112687285   0.019044390603128\\
	  -0.001106195440010   0.008512752503735\\
	  -0.000475634312656   0.004105807235686\\
	  -0.000175930557418   0.001892119140126\\
	  -0.000035771942102   0.000912683672872\\
	   0.000019591820855   0.000196242816461\\
	   0.000024793822889  -0.000091374359170\\
	   0.000008365703585  -0.000236584512966\\
	   0.000003722237800   0.000142601503432\\
	   0.000009349335294  -0.000029588032389\\
	   0.000006315293915  -0.000031058799994\\
	   0.000004564546274  -0.000004021758894\\
	   0.000004871932296   0.000010123261513\\
	   0.000006535239579   0.000023087092437\\
	   0.000008411547271   0.000014447222453\\
	   0.000006275342822  -0.000056932395299\\
	   0.000002075353411  -0.000027117673692\\
	   0.000003336839932   0.000052093060027\\
	   0.000005095464374  -0.000016727346196\\
	   0.000006183678493   0.000038312837629\\
	   0.000006496388768  -0.000031847254085\\
	   0.000003650494570  -0.000025063690404\\
	   0.000001736349115  -0.000013236751930\\
	   0.000000770096838  -0.000006100707325\\
	   0.000000333672037  -0.000002634139591\\
	   0.000000145424715  -0.000001133562456\\
	   0.000000064327365  -0.000000489564163\\
	   0.000000027882374  -0.000000239744782\\
	   0.000000010363088  -0.000000110864482\\
	   0.000000001604970  -0.000000064354841\\
	  -0.000000004903191  -0.000000065740869\\
	  -0.000000011054067  -0.000000057242826\\
	  -0.000000016350716  -0.000000048664860\\
	  -0.000000019650212  -0.000000017388577\\
	  -0.000000020138668   0.000000007548007\\
	  -0.000000018894468   0.000000017294115\\
	  -0.000000017218669   0.000000016208871\\
	  -0.000000015855835   0.000000011050328\\
	  -0.000000015008964   0.000000005894621\\
	  -0.000000014670637   0.000000000883940\\
	  -0.000000014616972   0.000000000190948\\
	}; \addlegendentry{$x^{3}$}
	\addplot [color=blue,solid,mark=star,mark options={solid}]
	  table[row sep=crcr]{%
	  -1.800000000000000   2.000000000000000\\
	  -1.457122036462741   4.854228151280546\\
	  -1.008944302000731   4.108702871069220\\
	  -0.659324107779626   2.883750669363618\\
	  -0.419072051374610   1.921425055884551\\
	  -0.262213150841450   1.215905189330313\\
	  -0.162767157422151   0.773107451089089\\
	  -0.099525532396835   0.491783822714175\\
	  -0.059328495268629   0.312194924026266\\
	  -0.034652137307722   0.181372712398730\\
	  -0.020603897124752   0.099621197322718\\
	  -0.012802320531968   0.056424661991637\\
	  -0.008347131371246   0.032686528227678\\
	  -0.005454107776973   0.025172169572091\\
	  -0.003227193099469   0.019364777694712\\
	  -0.001498319327123   0.015211356568790\\
	  -0.000508845845373   0.004584220888301\\
	  -0.000173604630362   0.002121724354433\\
	  -0.000025589925516   0.000839227385685\\
	   0.000026995512390   0.000212857301627\\
	   0.000035729932421  -0.000037984117908\\
	   0.000028524312169  -0.000106051544970\\
	   0.000018343186079  -0.000097549780350\\
	   0.000010118457863  -0.000066947440232\\
	   0.000004950239448  -0.000036428005782\\
	   0.000002194869776  -0.000018686519036\\
	   0.000000841218291  -0.000008391330210\\
	   0.000000277239654  -0.000002891259529\\
	   0.000000122684643  -0.000000201652824\\
	   0.000000131741727   0.000000382282705\\
	   0.000000167417867   0.000000331183586\\
	   0.000000170889241  -0.000000261272021\\
	   0.000000162793043   0.000000099070261\\
	   0.000000192651662   0.000000497686979\\
	   0.000000209429157  -0.000000161634225\\
	   0.000000187581254  -0.000000275168537\\
	   0.000000129168998  -0.000000892399935\\
	   0.000000047565975  -0.000000739561217\\
	  -0.000000007781401  -0.000000367542578\\
	  -0.000000030517533  -0.000000087350418\\
	  -0.000000032375310   0.000000050085515\\
	  -0.000000025445130   0.000000088466844\\
	  -0.000000017049782   0.000000079424332\\
	  -0.000000010284254   0.000000055887048\\
	  -0.000000005797608   0.000000033851772\\
	  -0.000000003209225   0.000000017921981\\
	  -0.000000001905491   0.000000008157211\\
	  -0.000000001335793   0.000000003239260\\
	  -0.000000001109121   0.000000001295177\\
	  -0.000000000998451   0.000000000918223\\
	  -0.000000000906670   0.000000000917140\\
	}; \addlegendentry{$x^{4}$}
	\end{axis}
	\end{tikzpicture}%
	
	\caption{State trajectory of the global plant $\boldsymbol{x}$.} \label{fig:SR.1}
\end{figure}
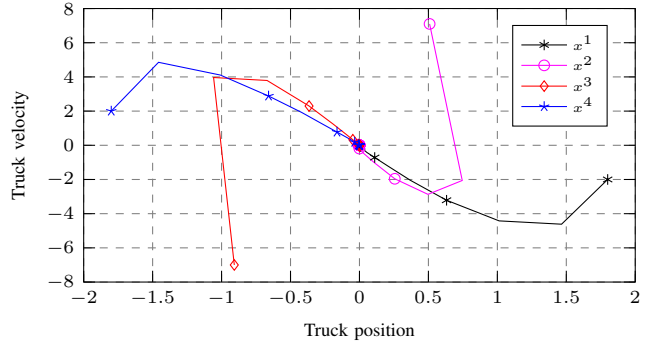

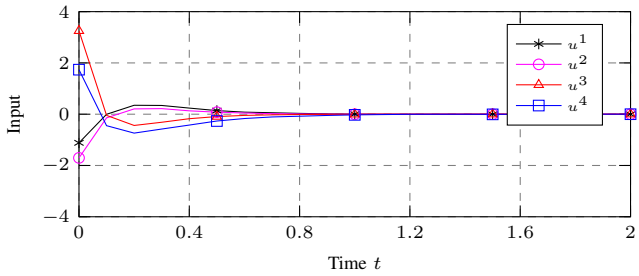
\begin{figure}[!t]
	\centering
	\pgfplotsset{major grid style={dashed,gray}}
	\pgfplotsset{minor grid style={dotted,gray}}
	\pgfplotsset{
		every axis legend/.append style={
		at={(0.95,0.94)},
		anchor=north east,
		row sep=-3pt,
	}}
	\setlength\figureheight{0.15\textwidth}
	\setlength\figurewidth{0.4\textwidth}
	
	\definecolor{mycolor1}{rgb}{1.00000,0.00000,1.00000}%
	\begin{tikzpicture}
	
	\begin{axis}[%
	width=\figurewidth,
	height=\figureheight,
	at={(0\figurewidth,0\figureheight)},
	scale only axis,
	xmin=0,
	xmax=2,
	xlabel={Time $t$},
	xtick={0,0.4,...,2},
	xticklabel style = {font=\scriptsize},
	xlabel style = {font=\scriptsize},
	xmajorgrids,
	xminorgrids,
	ymin=-4,
	ymax=4,
	ylabel={Input},
	ytick={-4,-2,...,4},
	yticklabel style = {font=\scriptsize},
	ylabel style = {font=\scriptsize},
	ymajorgrids,
	yminorgrids,
	legend style={legend plot pos=left},
	scaled ticks = false,
	mark repeat={5}
	]
	
	\addplot [color=black,solid,mark=asterisk,mark options={solid}]
	  table[row sep=crcr]{%
	   0.000000000000000  -1.116493977477002\\
	   0.100000000000000  -0.003698883850805\\
	   0.200000000000000   0.348195147116928\\
	   0.300000000000000   0.340972590547795\\
	   0.400000000000000   0.238273479276888\\
	   0.500000000000000   0.140512450172833\\
	   0.600000000000000   0.080050725461503\\
	   0.700000000000000   0.056754102705159\\
	   0.800000000000000   0.032261447774084\\
	   0.900000000000000   0.014004239528836\\
	   1.000000000000000   0.004664872690628\\
	   1.100000000000000   0.000956837695885\\
	   1.200000000000000  -0.000078689144257\\
	   1.300000000000000  -0.000172868650699\\
	   1.400000000000000  -0.000065830156362\\
	   1.500000000000000   0.000014819219130\\
	   1.600000000000000   0.000041757314798\\
	   1.700000000000000   0.000038836675568\\
	   1.800000000000000   0.000026534516564\\
	   1.900000000000000   0.000015155528933\\
	   2.000000000000000   0.000007586010356\\
	   2.100000000000000   0.000004499500353\\
	   2.200000000000000   0.000001302298451\\
	   2.300000000000000   0.000000711075807\\
	   2.400000000000000   0.000000482608767\\
	   2.500000000000000   0.000000142186486\\
	   2.600000000000000   0.000000110338773\\
	   2.700000000000000   0.000000098981442\\
	   2.800000000000000   0.000000068589141\\
	   2.900000000000000   0.000000043414680\\
	   3.000000000000000   0.000000052997348\\
	   3.100000000000001   0.000000042925071\\
	   3.200000000000000   0.000000058286292\\
	   3.300000000000000   0.000000054039879\\
	   3.400000000000000   0.000000048720182\\
	   3.500000000000000   0.000000045465397\\
	   3.600000000000001   0.000000043532483\\
	   3.700000000000000   0.000000045880120\\
	   3.800000000000000   0.000000045945193\\
	   3.900000000000000   0.000000045930190\\
	   4.000000000000000   0.000000045820806\\
	   4.100000000000001   0.000000046032521\\
	   4.200000000000000   0.000000045373124\\
	   4.300000000000001   0.000000044293128\\
	   4.400000000000000   0.000000046425500\\
	   4.500000000000000   0.000000046427648\\
	   4.600000000000001   0.000000046170995\\
	   4.700000000000000   0.000000045956759\\
	   4.800000000000001   0.000000045867215\\
	   4.900000000000000   0.000000045526842\\
	}; \addlegendentry{$u^{1}$}
	\addplot [color=mycolor1,solid,mark=o,mark options={solid}]
	  table[row sep=crcr]{%
	   0.000000000000000  -1.703740944433644\\
	   0.100000000000000  -0.143488400702465\\
	   0.200000000000000   0.206621341836858\\
	   0.300000000000000   0.217694852098757\\
	   0.400000000000000   0.131470388648586\\
	   0.500000000000000   0.079550900962355\\
	   0.600000000000000   0.045658189452088\\
	   0.700000000000000   0.021248187302594\\
	   0.800000000000000   0.010235992525932\\
	   0.900000000000000   0.003043127515393\\
	   1.000000000000000  -0.000277286980019\\
	   1.100000000000000  -0.001090511918539\\
	   1.200000000000000  -0.000926326954353\\
	   1.300000000000000  -0.000558411319596\\
	   1.400000000000000  -0.000275551877351\\
	   1.500000000000000  -0.000119938027072\\
	   1.600000000000000  -0.000044518131434\\
	   1.700000000000000  -0.000014867909995\\
	   1.800000000000000  -0.000005570222164\\
	   1.900000000000000  -0.000003619699680\\
	   2.000000000000000  -0.000002302416175\\
	   2.100000000000000  -0.000001621514519\\
	   2.200000000000000  -0.000001078434107\\
	   2.300000000000000  -0.000001068637473\\
	   2.400000000000000  -0.000000650036673\\
	   2.500000000000000  -0.000000315899128\\
	   2.600000000000000  -0.000000069056337\\
	   2.700000000000000  -0.000000125061477\\
	   2.800000000000000  -0.000000225442000\\
	   2.900000000000000  -0.000000096406255\\
	   3.000000000000000  -0.000000153139547\\
	   3.100000000000001   0.000000009515024\\
	   3.200000000000000  -0.000000008880294\\
	   3.300000000000000  -0.000000029607814\\
	   3.400000000000000  -0.000000030526048\\
	   3.500000000000000  -0.000000060924017\\
	   3.600000000000001  -0.000000045801430\\
	   3.700000000000000  -0.000000047258578\\
	   3.800000000000000  -0.000000047310477\\
	   3.900000000000000  -0.000000047884375\\
	   4.000000000000000  -0.000000046886316\\
	   4.100000000000001  -0.000000036341794\\
	   4.200000000000000  -0.000000061754115\\
	   4.300000000000001  -0.000000045938865\\
	   4.400000000000000  -0.000000046068920\\
	   4.500000000000000  -0.000000046964328\\
	   4.600000000000001  -0.000000047240971\\
	   4.700000000000000  -0.000000047472415\\
	   4.800000000000001  -0.000000036100459\\
	   4.900000000000000  -0.000000062220422\\
	}; \addlegendentry{$u^{2}$}
	\addplot [color=red,solid,mark=triangle,mark options={solid}]
	  table[row sep=crcr]{%
	   0.000000000000000   3.252773648296913\\
	   0.100000000000000  -0.046115164645030\\
	   0.200000000000000  -0.439036718980064\\
	   0.300000000000000  -0.314264818964430\\
	   0.400000000000000  -0.175580496088804\\
	   0.500000000000000  -0.092220151799956\\
	   0.600000000000000  -0.047348666583736\\
	   0.700000000000000  -0.025035290660265\\
	   0.800000000000000  -0.012260125606382\\
	   0.900000000000000  -0.006321169338305\\
	   1.000000000000000  -0.003237624533964\\
	   1.100000000000000  -0.001369982479678\\
	   1.200000000000000  -0.000686243419671\\
	   1.300000000000000  -0.000319159016427\\
	   1.400000000000000  -0.000242850366494\\
	   1.500000000000000  -0.000118565269589\\
	   1.600000000000000  -0.000053840385584\\
	   1.700000000000000   0.000108463711919\\
	   1.800000000000000  -0.000053697846925\\
	   1.900000000000000  -0.000001516873645\\
	   2.000000000000000   0.000007781622390\\
	   2.100000000000000   0.000004324477539\\
	   2.200000000000000   0.000004157783653\\
	   2.300000000000000  -0.000002277166512\\
	   2.400000000000000  -0.000021342618847\\
	   2.500000000000000   0.000008811957509\\
	   2.600000000000000   0.000023884988231\\
	   2.700000000000000  -0.000020459106856\\
	   2.800000000000000   0.000016661414165\\
	   2.900000000000000  -0.000020910416015\\
	   3.000000000000000   0.000001964167550\\
	   3.100000000000001   0.000003489533110\\
	   3.200000000000000   0.000002108658942\\
	   3.300000000000000   0.000001023525931\\
	   3.400000000000000   0.000000443909714\\
	   3.500000000000000   0.000000190982100\\
	   3.600000000000001   0.000000076829212\\
	   3.700000000000000   0.000000041474381\\
	   3.800000000000000   0.000000016464857\\
	   3.900000000000000   0.000000001478759\\
	   4.000000000000000   0.000000003963076\\
	   4.100000000000001   0.000000003742765\\
	   4.200000000000000   0.000000010383989\\
	   4.300000000000001   0.000000008788713\\
	   4.400000000000000   0.000000004342521\\
	   4.500000000000000   0.000000001152660\\
	   4.600000000000001  -0.000000000047510\\
	   4.700000000000000  -0.000000000049066\\
	   4.800000000000001  -0.000000000019765\\
	   4.900000000000000   0.000000001115426\\
	}; \addlegendentry{$u^{3}$}
	\addplot [color=blue,solid,mark=square,mark options={solid}]
	  table[row sep=crcr]{%
	   0.000000000000000   1.736376760677725\\
	   0.100000000000000  -0.447453348165078\\
	   0.200000000000000  -0.737397766114549\\
	   0.300000000000000  -0.578737744035832\\
	   0.400000000000000  -0.423753393668614\\
	   0.500000000000000  -0.265730757608579\\
	   0.600000000000000  -0.168675739545561\\
	   0.700000000000000  -0.107582408109378\\
	   0.800000000000000  -0.078338686113613\\
	   0.900000000000000  -0.048963945534984\\
	   1.000000000000000  -0.025879607223409\\
	   1.100000000000000  -0.014227469913725\\
	   1.200000000000000  -0.004497765545822\\
	   1.300000000000000  -0.003464467279508\\
	   1.400000000000000  -0.002465848228016\\
	   1.500000000000000  -0.006356532057375\\
	   1.600000000000000  -0.001470621537702\\
	   1.700000000000000  -0.000765320738691\\
	   1.800000000000000  -0.000374650287310\\
	   1.900000000000000  -0.000149913017127\\
	   2.000000000000000  -0.000040699657721\\
	   2.100000000000000   0.000004995673793\\
	   2.200000000000000   0.000018175331071\\
	   2.300000000000000   0.000018094800575\\
	   2.400000000000000   0.000010469020865\\
	   2.500000000000000   0.000006258869993\\
	   2.600000000000000   0.000003348856474\\
	   2.700000000000000   0.000001421255275\\
	   2.800000000000000   0.000000351083894\\
	   2.900000000000000  -0.000000204864428\\
	   3.000000000000000  -0.000000323050440\\
	   3.100000000000001   0.000000256286991\\
	   3.200000000000000   0.000000264156558\\
	   3.300000000000000  -0.000000382422894\\
	   3.400000000000000  -0.000000062215813\\
	   3.500000000000000  -0.000000368509578\\
	   3.600000000000001   0.000000090951497\\
	   3.700000000000000   0.000000222159093\\
	   3.800000000000000   0.000000167447330\\
	   3.900000000000000   0.000000082256924\\
	   4.000000000000000   0.000000023190657\\
	   4.100000000000001  -0.000000005103098\\
	   4.200000000000000  -0.000000013744612\\
	   4.300000000000001  -0.000000012916471\\
	   4.400000000000000  -0.000000009345187\\
	   4.500000000000000  -0.000000005707447\\
	   4.600000000000001  -0.000000002825916\\
	   4.700000000000000  -0.000000001046305\\
	   4.800000000000001  -0.000000000100953\\
	   4.900000000000000   0.000000000136633\\
	}; \addlegendentry{$u^{4}$}
	\end{axis}
	\end{tikzpicture}%

	\caption{Input sequence of the global plant $\boldsymbol{u}$.} \label{fig:SR.2}
\end{figure}

Figure \ref*{fig:SR.1} shows the state trajectory followed by all trucks when the control action is computed using Algorithm \ref*{alg:1}. Figure \ref*{fig:SR.2} depicts the input signal applied to each truck (following the disturbance rejection policy \Eqref{eq:DTube.5}). As expected, starting from a point inside the region of attraction of the controller: (i) the optimization is always feasible, (ii) global constraint satisfaction is attained, and (iii) the global state converges asymptotically to the origin.

Algorithm \ref*{alg:1} is compared to three different approaches: (i) a standard centralized MPC (CMPC), (ii) a decentralized scheme (TMPC) where all the coupling is treated as a disturbance (which amounts to use \Eqref{eq:DeTube.5} as a control policy), and (iii) a non-robust decentralized approach where the dynamic coupling is entirely neglected (DeMPC). All four control architectures are implemented using the same controller parameters (weight matrices in the cost function, and local linear controllers). Figures \ref*{fig:SR.3} shows the true state trajectory of the first truck under the action of all control schemes. The path followed by the truck under Algorithm \ref*{alg:1} lies closer to the CMPC trajectory than the result provided by the TMPC. This enhanced performance is expected, as Algorithm \ref*{alg:1} makes use of information that the TMPC does not posses. Figure \ref*{fig:SR.4} shows the true trajectory of truck number 3. Note how all distributed approaches have a lower impact over the third truck. This behaviour is expected due to the weak coupling between truck 3 and its neighbours.

\begin{figure}[ht!]
	\centering
	\pgfplotsset{major grid style={dashed,gray}}
	\pgfplotsset{minor grid style={dotted,gray}}
	\pgfplotsset{
		every axis legend/.append style={
		at={(0.85,0.94)},
		anchor=north east,
	}}
	\setlength\figureheight{0.2\textwidth}
	\setlength\figurewidth{0.4\textwidth}

	\definecolor{mycolor1}{rgb}{1,0,1}%
	\begin{tikzpicture}
	
	\begin{axis}[%
	width=\figurewidth,
	height=\figureheight,
	at={(0\figurewidth,0\figureheight)},
	scale only axis,
	xmin=-0.5,
	xmax=2,
	xlabel={Truck position},
	xtick={-0.5,0.0,0.5,1.0,1.5,2},
	xticklabel style = {font=\scriptsize},
	xlabel style = {font=\scriptsize},
	xmajorgrids,
	xminorgrids,
	ymin=-6,
	ymax=1,
	ylabel={Truck velocity},
	ytick={-6,-5,...,1},
	yticklabel style = {font=\scriptsize},
	ylabel style = {font=\scriptsize},
	ymajorgrids,
	yminorgrids,
	legend style={legend plot pos=left},
	scaled ticks = false,
	]
	\addplot [color=mycolor1,only marks,mark=asterisk,mark options={solid},forget plot]
	  table[row sep=crcr]{%
	1.8	-2\\
	};
	\addplot [color=blue,only marks,mark=asterisk,mark options={solid},forget plot]
	  table[row sep=crcr]{%
	1.8	-2\\
	};
	\addplot [color=red,only marks,mark=asterisk,mark options={solid},forget plot]
	  table[row sep=crcr]{%
	1.8	-2\\
	};
	\addplot [color=black,only marks,mark=asterisk,mark options={solid},forget plot]
	  table[row sep=crcr]{%
	1.8	-2\\
	}; \label{initial_state}
	
	\addplot [color=mycolor1,dash pattern=on 4pt off 1pt on 4pt off 4pt]
	  table[row sep=crcr]{%
	   1.800000000000000  -2.000000000000000\\
	   1.487128564252778  -4.195574092261435\\
	   1.058007114631509  -4.365201174626686\\
	   0.675092098761212  -3.300808313650090\\
	   0.402684861130537  -2.161320331071472\\
	   0.230349970215231  -1.297404447563909\\
	   0.129128734596849  -0.735300123792146\\
	   0.072497638044771  -0.402364439838589\\
	   0.041644076573388  -0.217530693086822\\
	   0.024879773151580  -0.119242142106169\\
	   0.015555144064342  -0.068001166765508\\
	   0.010121823482898  -0.041038325419743\\
	   0.006768365847042  -0.026220544021392\\
	   0.004588047853429  -0.017489094786249\\
	   0.003119242173480  -0.011948990777961\\
	   0.002112834571238  -0.008220130648565\\
	   0.001421492207226  -0.005635377061808\\
	   0.000949246833770  -0.003829992363577\\
	   0.000629629977621  -0.002576887523011\\
	   0.000415393237226  -0.001718015968532\\
	   0.000272999992273  -0.001136837187422\\
	   0.000178991139929  -0.000748066346166\\
	   0.000117222042581  -0.000490469038760\\
	   0.000076756633305  -0.000320922607819\\
	   0.000050284071142  -0.000209894104924\\
	   0.000032964922765  -0.000137379447187\\
	   0.000021624377206  -0.000090012058548\\
	   0.000014190534616  -0.000059043570888\\
	   0.000009312372088  -0.000038767225966\\
	   0.000006108568849  -0.000025470962217\\
	   0.000004003646640  -0.000016734024790\\
	   0.000002619726503  -0.000011013890070\\
	   0.000001708313749  -0.000007259852779\\
	   0.000001104766362  -0.000004839753018\\
	   0.000000700330529  -0.000003267077256\\
	   0.000000425410748  -0.000002242639512\\
	   0.000000248859924  -0.000001301935547\\
	   0.000000117855423  -0.000001312461278\\
	   0.000000031173474  -0.000000437029997\\
	  -0.000000017641993  -0.000000535065383\\
	  -0.000000069115693  -0.000000493194580\\
	  -0.000000121727655  -0.000000555467640\\
	  -0.000000145865382   0.000000060033792\\
	  -0.000000147405603  -0.000000087493866\\
	  -0.000000156503951  -0.000000093950402\\
	  -0.000000171579933  -0.000000204483396\\
	  -0.000000191793910  -0.000000199074052\\
	  -0.000000216915477  -0.000000300064988\\
	  -0.000000245559756  -0.000000272245476\\
	  -0.000000277064195  -0.000000354697228\\
	  -0.000000268645674   0.000000503632600\\
	}; \addlegendentry{DeMPC}
	\addplot [color=blue,densely dotted]
	  table[row sep=crcr]{%
	   1.800000000000000  -2.000000000000000\\
	   1.463821852156324  -4.650619550900178\\
	   0.988190413494766  -4.838022124501427\\
	   0.578329432794714  -3.374654642435008\\
	   0.308302606277052  -2.044210720655241\\
	   0.147059829376687  -1.192854745642975\\
	   0.059422520637772  -0.570086171762622\\
	   0.020221333757969  -0.220085947680225\\
	   0.005991609162228  -0.067312360159289\\
	   0.001942231637033  -0.014676754477537\\
	   0.001113572441759  -0.002140628051768\\
	   0.000940578273639  -0.001330068465314\\
	   0.000765964570380  -0.002136987976733\\
	   0.000546074412434  -0.002249157984943\\
	   0.000344830174480  -0.001777820385152\\
	   0.000197851978307  -0.001169151942165\\
	   0.000105778420072  -0.000679372204151\\
	   0.000054066321198  -0.000359819822317\\
	   0.000027211740647  -0.000180147574830\\
	   0.000013883585555  -0.000087910924888\\
	   0.000007259401312  -0.000045245352279\\
	   0.000004188050243  -0.000016688615816\\
	   0.000002699341901  -0.000013103232041\\
	   0.000001549347593  -0.000009919519114\\
	   0.000000750065953  -0.000006117344452\\
	   0.000000280093964  -0.000003324605006\\
	   0.000000047339005  -0.000001364378877\\
	  -0.000000043224444  -0.000000463154149\\
	  -0.000000091583973  -0.000000501175820\\
	  -0.000000150354131  -0.000000668067596\\
	  -0.000000224459556  -0.000000807844976\\
	  -0.000000308591316  -0.000000869904470\\
	  -0.000000401192277  -0.000000975899201\\
	  -0.000000483545857  -0.000000674443218\\
	  -0.000000535624334  -0.000000371687644\\
	  -0.000000565990277  -0.000000237352632\\
	  -0.000000518490242   0.000001154428834\\
	  -0.000000408533550   0.000001042609136\\
	  -0.000000325545828   0.000000623016359\\
	  -0.000000286413727   0.000000168109438\\
	  -0.000000287006670  -0.000000172373234\\
	  -0.000000331052796  -0.000000695091614\\
	  -0.000000415991519  -0.000000993508260\\
	  -0.000000497089697  -0.000000633089420\\
	  -0.000000548703713  -0.000000402174268\\
	  -0.000000513087465   0.000001080058844\\
	  -0.000000408458295   0.000001009728230\\
	  -0.000000327767864   0.000000609614072\\
	  -0.000000289645424   0.000000161215873\\
	  -0.000000290565997  -0.000000172176283\\
	  -0.000000334891727  -0.000000700739039\\
	}; \addlegendentry{TMPC}
	\addplot [color=red,densely dashed]
	  table[row sep=crcr]{%
	   1.800000000000000  -2.000000000000000\\
	   1.465615006851268  -4.615609597664649\\
	   1.013007538158220  -4.421983967822482\\
	   0.631730368991285  -3.214517949674889\\
	   0.369349807922522  -2.048093893198891\\
	   0.206719576872104  -1.216229460490498\\
	   0.110482426183950  -0.715633520190633\\
	   0.053963841286344  -0.418978821664114\\
	   0.022926594706335  -0.205225364438006\\
	   0.008603307854417  -0.083353221955169\\
	   0.003056273871331  -0.028574267544019\\
	   0.001225794949702  -0.008407424578358\\
	   0.000679494696744  -0.002624396016333\\
	   0.000476784639522  -0.001447524607599\\
	   0.000342579540825  -0.001235694675596\\
	   0.000229520228288  -0.001025454713129\\
	   0.000141342673600  -0.000740751999862\\
	   0.000080784322009  -0.000473826331989\\
	   0.000043509079788  -0.000274555803454\\
	   0.000022494654211  -0.000147678022632\\
	   0.000011395374259  -0.000075450958915\\
	   0.000005741370364  -0.000038221277848\\
	   0.000003010915686  -0.000016751397837\\
	   0.000001731181595  -0.000008963121114\\
	   0.000001058438886  -0.000004561563683\\
	   0.000000733159947  -0.000001987688144\\
	   0.000000578665816  -0.000001115156445\\
	   0.000000489719808  -0.000000669955858\\
	   0.000000438660660  -0.000000356078170\\
	   0.000000414684384  -0.000000127529802\\
	   0.000000402204408  -0.000000121679734\\
	   0.000000394805381  -0.000000028073719\\
	   0.000000390100513  -0.000000065006421\\
	   0.000000385152464  -0.000000034428247\\
	   0.000000382991236  -0.000000009265708\\
	   0.000000382609012   0.000000001399995\\
	   0.000000383117323   0.000000008585300\\
	   0.000000383499800  -0.000000000744279\\
	   0.000000383426998  -0.000000000709488\\
	   0.000000383367486  -0.000000000483301\\
	   0.000000383330349  -0.000000000262794\\
	   0.000000383283305  -0.000000000667145\\
	   0.000000383249698  -0.000000000018021\\
	   0.000000383516408   0.000000005224547\\
	   0.000000383627109  -0.000000002835931\\
	   0.000000383375100  -0.000000002207706\\
	   0.000000383225606  -0.000000000807100\\
	   0.000000383181799  -0.000000000083305\\
	   0.000000383179401   0.000000000032879\\
	   0.000000383174682  -0.000000000123587\\
	   0.000000383457867   0.000000005647170\\
	}; \addlegendentry{Algorithm 1}
	\addplot [color=black,solid]
	  table[row sep=crcr]{%
	   1.800000000000000  -2.000000000000000\\
	   1.444619274210953  -5.025534976638292\\
	   0.985895473197174  -4.149307437528706\\
	   0.636415869764775  -2.854509964661423\\
	   0.401948628095989  -1.847455241739187\\
	   0.251635663659694  -1.167655040353653\\
	   0.156991255727053  -0.730997833446819\\
	   0.097826763592357  -0.455942157634315\\
	   0.060942745849344  -0.284023642186665\\
	   0.037968874402095  -0.176878704724974\\
	   0.023660717538772  -0.110171369241889\\
	   0.014747937430941  -0.068636319378506\\
	   0.009194634333825  -0.042773368115617\\
	   0.005733601603494  -0.026661297807303\\
	   0.003576048280917  -0.016623051751232\\
	   0.002230794334179  -0.010365108313262\\
	   0.001391814433568  -0.006466216583322\\
	   0.000868405693955  -0.004034220284610\\
	   0.000541975230534  -0.002514574135104\\
	   0.000338417401142  -0.001569121208558\\
	   0.000211026423155  -0.000986347245223\\
	   0.000131229122104  -0.000614540303593\\
	   0.000081517077468  -0.000382781957100\\
	   0.000050574940930  -0.000237990675501\\
	   0.000031341737530  -0.000147875546877\\
	   0.000019414210499  -0.000091432908370\\
	   0.000012018482587  -0.000056940370104\\
	   0.000007463543157  -0.000034468255138\\
	   0.000004815701629  -0.000018728202289\\
	   0.000003183978653  -0.000013944619532\\
	   0.000002035153815  -0.000009091899916\\
	   0.000001289404202  -0.000005863790024\\
	   0.000000811272523  -0.000003726441173\\
	   0.000000511333470  -0.000002291740283\\
	   0.000000321417150  -0.000001515921670\\
	   0.000000188063195  -0.000001153655038\\
	   0.000000082500799  -0.000000957554063\\
	   0.000000020622264  -0.000000292230944\\
	  -0.000000008076154  -0.000000280795756\\
	  -0.000000027773415  -0.000000115992912\\
	  -0.000000028190317   0.000000102807412\\
	  -0.000000020438521   0.000000053012624\\
	  -0.000000016756046   0.000000021190948\\
	  -0.000000025558453  -0.000000192129271\\
	  -0.000000043174675  -0.000000160171630\\
	  -0.000000058498151  -0.000000145974992\\
	  -0.000000072060826  -0.000000125175781\\
	  -0.000000075939276   0.000000044006768\\
	  -0.000000072682558   0.000000021492481\\
	  -0.000000050827131   0.000000406152698\\
	}; \addlegendentry{CMPC}
	
	\addplot [color=mycolor1,only marks,mark=o,mark options={solid},forget plot]
	  table[row sep=crcr]{%
	  -0.000000268645674   0.000000503632600\\
	};
	\addplot [color=blue,only marks,mark=o,mark options={solid},forget plot]
	  table[row sep=crcr]{%
	  -0.000000334891727  -0.000000700739039\\
	};
	\addplot [color=red,only marks,mark=o,mark options={solid},forget plot]
	  table[row sep=crcr]{%
	   0.000000383457867   0.000000005647170\\
	};
	\addplot [color=black,only marks,mark=o,mark options={solid},forget plot]
	  table[row sep=crcr]{%
	  -0.000000050827131   0.000000406152698\\
	}; \label{final_state}
	\end{axis}
	\end{tikzpicture}%

	\caption{State trajectory of truck 1: \ref*{initial_state} $x^{1}_{t=0}$, \ref*{final_state} $x^{1}_{t=50}$.} \label{fig:SR.3}
\end{figure}

\begin{figure}[ht!]
	\centering
	\pgfplotsset{major grid style={dashed,gray}}
	\pgfplotsset{minor grid style={dotted,gray}}
	\pgfplotsset{
		every axis legend/.append style={
		at={(0.95,0.5)},
		anchor=north east,
	}}
	\setlength\figureheight{0.2\textwidth}
	\setlength\figurewidth{0.4\textwidth}
	
	\definecolor{mycolor1}{rgb}{1,0,1}%
	\begin{tikzpicture}
	
	\begin{axis}[%
	width=\figurewidth,
	height=\figureheight,
	at={(0\figurewidth,0\figureheight)},
	scale only axis,
	xmin=-1.2,
	xmax=0.2,
	xlabel={Truck position},
	xtick={-1.2,-1.0,-0.8,-0.6,-0.4,-0.2,-0.0,0.2},
	xticklabel style = {font=\scriptsize},
	xlabel style = {font=\scriptsize},
	xmajorgrids,
	xminorgrids,
	ymin=-8,
	ymax=6,
	ylabel={Truck velocity},
	ytick={-8,-6,...,6},
	yticklabel style = {font=\scriptsize},
	ylabel style = {font=\scriptsize},
	ymajorgrids,
	yminorgrids,
	legend style={legend plot pos=left},
	scaled ticks = false,
	]
	\addplot [color=mycolor1,only marks,mark=asterisk,mark options={solid},forget plot]
	table[row sep=crcr]{%
		-0.9077	-7\\
	};
	\addplot [color=blue,only marks,mark=asterisk,mark options={solid},forget plot]
	table[row sep=crcr]{%
		-0.9077	-7\\
	};
	\addplot [color=red,only marks,mark=asterisk,mark options={solid},forget plot]
	table[row sep=crcr]{%
		-0.9077	-7\\
	};
	\addplot [color=black,only marks,mark=asterisk,mark options={solid},forget plot]
	table[row sep=crcr]{%
		-0.9077	-7\\
	};
	
	\addplot [color=mycolor1,dash pattern=on 4pt off 1pt on 4pt off 4pt]
	  table[row sep=crcr]{%
	  -0.907700000000000  -7.000000000000000\\
	  -1.058728540779639   3.947440820236476\\
	  -0.672515437047517   3.773597349437412\\
	  -0.368488172126023   2.308469482681925\\
	  -0.190759394911559   1.247620081142230\\
	  -0.096315860366357   0.642182621805930\\
	  -0.048033199719814   0.323973622280818\\
	  -0.023751345669789   0.161922612893920\\
	  -0.011617783468849   0.080878337187723\\
	  -0.005567196363530   0.040198896145526\\
	  -0.002571069393542   0.019756984766487\\
	  -0.001112213021049   0.009437495785499\\
	  -0.000409498254503   0.004624675587415\\
	  -0.000076173661541   0.002046454704229\\
	   0.000061909025997   0.000717916641549\\
	   0.000102316201719   0.000091747644888\\
	   0.000108458756726   0.000031228684476\\
	   0.000108436013979  -0.000031491636658\\
	   0.000102775111479  -0.000081518483064\\
	   0.000092980029205  -0.000114188081106\\
	   0.000080359012253  -0.000138036657946\\
	   0.000067590772990  -0.000117267453711\\
	   0.000055954977872  -0.000115341199394\\
	   0.000043935666403  -0.000124898955084\\
	   0.000031788320472  -0.000117951087140\\
	   0.000020611186883  -0.000105520896408\\
	   0.000010813161902  -0.000090390576187\\
	   0.000002584257684  -0.000074157109228\\
	  -0.000004021639419  -0.000057946145798\\
	  -0.000009048432017  -0.000042587767330\\
	  -0.000012859313931  -0.000033620223331\\
	  -0.000015904234274  -0.000027267990685\\
	  -0.000018384527193  -0.000022328846207\\
	  -0.000020410400157  -0.000018181589842\\
	  -0.000022042727097  -0.000014460449955\\
	  -0.000023316162307  -0.000011006428434\\
	  -0.000024252898260  -0.000007729198301\\
	  -0.000024910912115  -0.000005431702297\\
	  -0.000025366100371  -0.000003673011774\\
	  -0.000025678864327  -0.000002582558876\\
	  -0.000025903071318  -0.000001901482411\\
	  -0.000026069433270  -0.000001425593457\\
	  -0.000026192208128  -0.000001029919189\\
	  -0.000026276310183  -0.000000652457739\\
	  -0.000026323317732  -0.000000288348956\\
	  -0.000026333743680   0.000000078809302\\
	  -0.000026308202365   0.000000430693248\\
	  -0.000026246728698   0.000000797062691\\
	  -0.000026145803771   0.000001219164555\\
	  -0.000026004014702   0.000001614031425\\
	  -0.000025822689091   0.000002009509104\\
	}; \addlegendentry{DeMPC}
	\addplot [color=blue,densely dotted]
	  table[row sep=crcr]{%
	  -0.907700000000000  -7.000000000000000\\
	  -1.057236320274230   3.977179834940952\\
	  -0.668710601006236   3.790128524538361\\
	  -0.364512568782152   2.295442327422630\\
	  -0.188211519249050   1.232114140229859\\
	  -0.094727556242126   0.638470418747529\\
	  -0.046488927207725   0.326785682207345\\
	  -0.021892708321293   0.165392781822236\\
	  -0.009714953185402   0.078310051583810\\
	  -0.004057766015772   0.034911318624665\\
	  -0.001570414265639   0.014872772032919\\
	  -0.000567053211820   0.005214211094287\\
	  -0.000231260462457   0.001509700418270\\
	  -0.000095968736209   0.001195778038379\\
	  -0.000005865187931   0.000607215565214\\
	   0.000035489710235   0.000220661659779\\
	   0.000027684101876  -0.000374877884271\\
	   0.000013049012606   0.000080925402157\\
	   0.000016904098396  -0.000003602840255\\
	   0.000013617658832  -0.000061915784504\\
	   0.000008418894949  -0.000042069597569\\
	   0.000005416878360  -0.000018015059124\\
	   0.000004151941645  -0.000007304110146\\
	   0.000003473669428  -0.000006257936519\\
	   0.000003144218873  -0.000000345939135\\
	   0.000001969198786  -0.000023073579080\\
	  -0.000000206068170  -0.000020418723268\\
	  -0.000001662737890  -0.000008736211072\\
	  -0.000002092335695   0.000000121362091\\
	  -0.000001868794772   0.000004334407220\\
	  -0.000001399006357   0.000005054592292\\
	  -0.000000936468827   0.000004194290207\\
	  -0.000000581193237   0.000002911687008\\
	  -0.000000346907478   0.000001775223372\\
	  -0.000000210205647   0.000000959975593\\
	  -0.000000139483022   0.000000455331522\\
	  -0.000000107441583   0.000000186008800\\
	  -0.000000094873518   0.000000065598283\\
	  -0.000000090019668   0.000000031535630\\
	  -0.000000087051318   0.000000027813871\\
	  -0.000000084320792   0.000000026773281\\
	  -0.000000081981843   0.000000020003657\\
	  -0.000000080464276   0.000000010362382\\
	  -0.000000079838437   0.000000002173353\\
	  -0.000000080251605  -0.000000010394289\\
	  -0.000000081627336  -0.000000017086505\\
	  -0.000000083299427  -0.000000016341336\\
	  -0.000000084576218  -0.000000009203885\\
	  -0.000000085129177  -0.000000001872333\\
	  -0.000000084978940   0.000000004855064\\
	  -0.000000084238965   0.000000009921743\\
	}; \addlegendentry{TMPC}
	\addplot [color=red,densely dashed]
	  table[row sep=crcr]{%
	  -0.907700000000000  -7.000000000000000\\
	  -1.057864895307967   3.964652730465833\\
	  -0.669891635001367   3.791569134218735\\
	  -0.365506502040563   2.297739296128228\\
	  -0.189174348214153   1.230451057402783\\
	  -0.095919256459080   0.635562107462069\\
	  -0.047987705990931   0.323556388891050\\
	  -0.023623851786343   0.163971515240130\\
	  -0.011418271577798   0.080277193646246\\
	  -0.005420644036320   0.039740925785791\\
	  -0.002483112687285   0.019044390603128\\
	  -0.001106195440010   0.008512752503735\\
	  -0.000475634312656   0.004105807235686\\
	  -0.000175930557418   0.001892119140126\\
	  -0.000035771942102   0.000912683672872\\
	   0.000019591820855   0.000196242816461\\
	   0.000024793822889  -0.000091374359170\\
	   0.000008365703585  -0.000236584512966\\
	   0.000003722237800   0.000142601503432\\
	   0.000009349335294  -0.000029588032389\\
	   0.000006315293915  -0.000031058799994\\
	   0.000004564546274  -0.000004021758894\\
	   0.000004871932296   0.000010123261513\\
	   0.000006535239579   0.000023087092437\\
	   0.000008411547271   0.000014447222453\\
	   0.000006275342822  -0.000056932395299\\
	   0.000002075353411  -0.000027117673692\\
	   0.000003336839932   0.000052093060027\\
	   0.000005095464374  -0.000016727346196\\
	   0.000006183678493   0.000038312837629\\
	   0.000006496388768  -0.000031847254085\\
	   0.000003650494570  -0.000025063690404\\
	   0.000001736349115  -0.000013236751930\\
	   0.000000770096838  -0.000006100707325\\
	   0.000000333672037  -0.000002634139591\\
	   0.000000145424715  -0.000001133562456\\
	   0.000000064327365  -0.000000489564163\\
	   0.000000027882374  -0.000000239744782\\
	   0.000000010363088  -0.000000110864482\\
	   0.000000001604970  -0.000000064354841\\
	  -0.000000004903191  -0.000000065740869\\
	  -0.000000011054067  -0.000000057242826\\
	  -0.000000016350716  -0.000000048664860\\
	  -0.000000019650212  -0.000000017388577\\
	  -0.000000020138668   0.000000007548007\\
	  -0.000000018894468   0.000000017294115\\
	  -0.000000017218669   0.000000016208871\\
	  -0.000000015855835   0.000000011050328\\
	  -0.000000015008964   0.000000005894621\\
	  -0.000000014670637   0.000000000883940\\
	  -0.000000014616972   0.000000000190948\\
	}; \addlegendentry{Algorithm 1}
	\addplot [color=black,solid]
	  table[row sep=crcr]{%
	  -0.907700000000000  -7.000000000000000\\
	  -1.062557715531110   3.871127779484633\\
	  -0.676081665797570   3.854685713983864\\
	  -0.364420010660465   2.380023829631133\\
	  -0.181965798217443   1.270676708289911\\
	  -0.086812095882503   0.633419690204922\\
	  -0.040024085389121   0.302895719656576\\
	  -0.017857463804959   0.140716823683746\\
	  -0.007669498856569   0.063180378214334\\
	  -0.003122823525992   0.027816995747953\\
	  -0.001150610964353   0.011657465108891\\
	  -0.000366915026667   0.004032137692988\\
	  -0.000089439327397   0.001522348613251\\
	   0.000009765416188   0.000464016084146\\
	   0.000035771669690   0.000057099714569\\
	   0.000035029371758  -0.000071545299526\\
	   0.000026771483242  -0.000093465188320\\
	   0.000025750205814   0.000072542229063\\
	   0.000022993368224  -0.000127042219259\\
	   0.000015345502378  -0.000026149079418\\
	   0.000012975867496  -0.000021235594253\\
	   0.000011239530050  -0.000013497919430\\
	   0.000005906433873  -0.000092869579499\\
	   0.000001126090928  -0.000002965558196\\
	  -0.000000885134380  -0.000037134966341\\
	  -0.000002387727410   0.000006962946543\\
	  -0.000002699456245  -0.000013133076158\\
	  -0.000003017027085   0.000006724061094\\
	  -0.000003089014371  -0.000008117748338\\
	  -0.000003221019159   0.000005437509409\\
	  -0.000002815894393   0.000002669506225\\
	  -0.000002266283341   0.000008300162851\\
	  -0.000001377863928   0.000009456054634\\
	  -0.000000952526440  -0.000000921724604\\
	  -0.000000939801438   0.000001169709230\\
	  -0.000000715400677   0.000003309584366\\
	  -0.000000564726642  -0.000000286578338\\
	  -0.000000415347162   0.000003261870971\\
	  -0.000000280809181  -0.000000560737075\\
	  -0.000000157856810   0.000003007683450\\
	   0.000000205385756   0.000004249839812\\
	   0.000000426326000   0.000000179257085\\
	   0.000000240864893  -0.000003874287791\\
	   0.000000083094969   0.000000706424061\\
	   0.000000281789463   0.000003257736751\\
	   0.000000399627336  -0.000000889450698\\
	   0.000000106273245  -0.000004962331539\\
	  -0.000000099536967   0.000000830424989\\
	  -0.000000218896375  -0.000003204122500\\
	  -0.000000295989934   0.000001648171659\\
	}; \addlegendentry{CMPC}
	
	\addplot [color=mycolor1,only marks,mark=o,mark options={solid},forget plot]
	table[row sep=crcr]{%
	  -0.000025822689091   0.000002009509104\\
	};
	\addplot [color=blue,only marks,mark=o,mark options={solid},forget plot]
	table[row sep=crcr]{%
	  -0.000000084238965   0.000000009921743\\
	};
	\addplot [color=red,only marks,mark=o,mark options={solid},forget plot]
	table[row sep=crcr]{%
	  -0.000000014616972   0.000000000190948\\
	};
	\addplot [color=black,only marks,mark=o,mark options={solid},forget plot]
	table[row sep=crcr]{%
	  -0.000000295989934   0.000001648171659\\
	};
	\end{axis}
	\end{tikzpicture}%
	
	\caption{State trajectory of truck 3: \ref*{initial_state} $x^{3}_{t=0}$, \ref*{final_state} $x^{3}_{t=50}$.} \label{fig:SR.4}
\end{figure}

Table \ref*{table:SR.1} shows the aggregated cost of the true trajectories followed by the trucks after 50 time instants. TMPC produces trajectories with higher costs than the alternative approaches, including the decentralized controller. This is expected, given that TMPC induces a high, and unnecessary, degree of conservativeness to the selection of the control action. The costs obtained for truck 3 confirm what is shown in figure \ref*{fig:SR.3}, that the distributed approaches have a small impact when the coupling between neighbouring subsystems is weak. Truck 1 on the other hand, given its stronger coupling with truck 2, benefits more from the distributed framework. Algorithm \ref{alg:1} performs closer to the centralized controller than any of the other approaches.

\setlength{\tabcolsep}{3pt}
\begin{table}[!t] 
\caption{Cost comparison between different controllers.} \label{table:SR.1}
\centering
\begin{tabular}{c|| >{$}c<{$} >{$}c<{$} >{$}c<{$}}
\hline
Controller & \text{Global plant} & \text{Truck }i=1 & \text{Truck }i=3\\
\hline
TMPC        & 6.3637 & 4.7977 & 0.0600\\
DeMPC       & 6.3042 & 4.7702 & 0.0600\\
Algorithm 1 & 6.2798 & 4.7509 & 0.0600\\
CMPC        & 6.2472 & 4.7215 & 0.0600\\
\hline
\end{tabular}
\end{table}

\section{Conclusions and future work} \label{sec:CFW}
In this paper, a new approach to the problem of distributed control of dynamically coupled system has been devised. Each time step, the optimal control action is obtained, for each agent, through a chain of two robust controllers. The inner controller defines a reference trajectory from a decentralized framework, and the outer one uses this information (shared amongst neighbours) to reduce uncertainty about the dynamical coupling and improve performance. Under standard assumptions in the MPC context, and trivial design choices, the controller proposed in Algorithm \ref*{alg:1} has recursive feasibility and provides a guarantee of asymptotic stability of the origin and robust constraint satisfaction. Future work will include a comprehensive approach for the computation of the input tracking set in \Eqref{eq:DTube.27} and the inclusion of coupled constraints.

\bibliographystyle{IEEEtran}

\end{document}